\documentclass[10pt]{iopart}
\usepackage{graphicx}
\usepackage{bm}
\usepackage{iopams} 
\usepackage{datetime}

\begin{document}

\title{Hybrid nanowire ion-to-electron transducers for integrated bioelectronic circuitry}

\author{D.J. Carrad$^{1,2}$, A.B. Mostert$^{3}$, A.R. Ullah$^{1}$, A.M. Burke$^{1,4}$, H.J. Joyce$^{5,6}$, H.H. Tan$^{5}$, C. Jagadish$^{5}$, P. Krogstrup$^{7}$, J. Nyg{\aa}rd$^{7}$, P. Meredith$^{3,8}$ and A.P. Micolich$^{1}$}
\address{$^{1}$ School of Physics, University of New South Wales, Sydney NSW 2052, Australia}
\address{$^{2}$ Walter Schottky Institut, Technische Universit\"{a}t M\"{u}nchen, Am Coulombwall 4, Garching 85748, Germany}
\address{$^{3}$ Centre for Organic Photonics and Electronics, School of Mathematics and Physics, University of Queensland, Brisbane QLD 4072, Australia}
\address{$^{4}$ Solid State Physics/NanoLund, Lund University, SE-221 00 Lund, Sweden}
\address{$^{5}$ Department of Electronic Materials Engineering, Research School of Physics and Engineering, The Australian National University, Canberra ACT 0200, Australia}
\address{$^{6}$ Department of Engineering, University of Cambridge, Cambridge CB3 0FA, United Kingdom}
\address{$^{7}$ Center for Quantum Devices, Niels Bohr Institute, University of Copenhagen, Universitetsparken 5, DK-2100 Copenhagen, Denmark}
\address{$^{8}$ Physics Department, Swansea University, Swansea SA2 8PP, Wales, United Kingdom}
\ead{adam.micolich@nanoelectronics.physics.unsw.edu.au}

\date{\today}

\begin{abstract}
A key task in the emerging field of bioelectronics is the transduction between ionic/protonic and electronic signals at high fidelity. This is a considerable challenge since the two carrier types exhibit intrinsically different physics and are best supported by very different materials types -- electronic signals in inorganic semiconductors and ionic/protonic signals in organic or bio-organic polymers, gels or electrolytes. Here we demonstrate a new class of organic-inorganic transducing interface featuring semiconducting nanowires electrostatically gated using a solid proton-transporting hygroscopic polymer. This model platform allows us to study the basic transducing mechanisms as well as deliver high fidelity signal conversion by tapping into and drawing together the best candidates from traditionally disparate realms of electronic materials research. By combining complementary $n-$ and $p-$type transducers we demonstrate functional logic with significant potential for scaling towards high-density integrated bioelectronic circuitry.

\end{abstract}

\maketitle

Ion-to-electron signal transduction is at the heart of modern bioelectronics. To date, the predominant means to affect such transduction has been via organic semiconductor-based devices seeking to exploit the biocompatibility and processing flexibility of these materials. Typically these devices are transistors where changes in ionic environment affect the electric current through a semiconducting channel via an electrostatic field (organic field-effect transistors) or by ions entering the semiconductor and chemically doping or de-doping it (organic electrochemical transistors).~\cite{BerggrenAdvMat07, OwensMRS10} The latter have shown particular promise for applications such as \emph{in vivo} recording of brain activity.~\cite{KhodagholyNatComm13} `Protonic' transistors featuring thin proton conductive films bridging PdH$_x$ electrodes~\cite{ZhongNatComm11, OrdinarioNatChem14} have also been developed towards this application. A key long-term goal is full-scale integration of these devices with conventional electronics to facilitate signal recording, data transmission and control of multiplexed transducer arrays.~\cite{StrakosasJAPS15} An important first step is the development of ionically-active amplifiers and logic-gates featuring complementary $n$- and $p$-type transistors. Fully ionic junction transistors~\cite{TybrandtPNAS10} and logic gates~\cite{TybrandtNatComm12} have been developed, as have complementary protonic devices,~\cite{DengSciRep13} but they face difficulties with slow switching, poorly defined off-states and/or low gain.

\begin{figure}
\includegraphics[width=12cm]{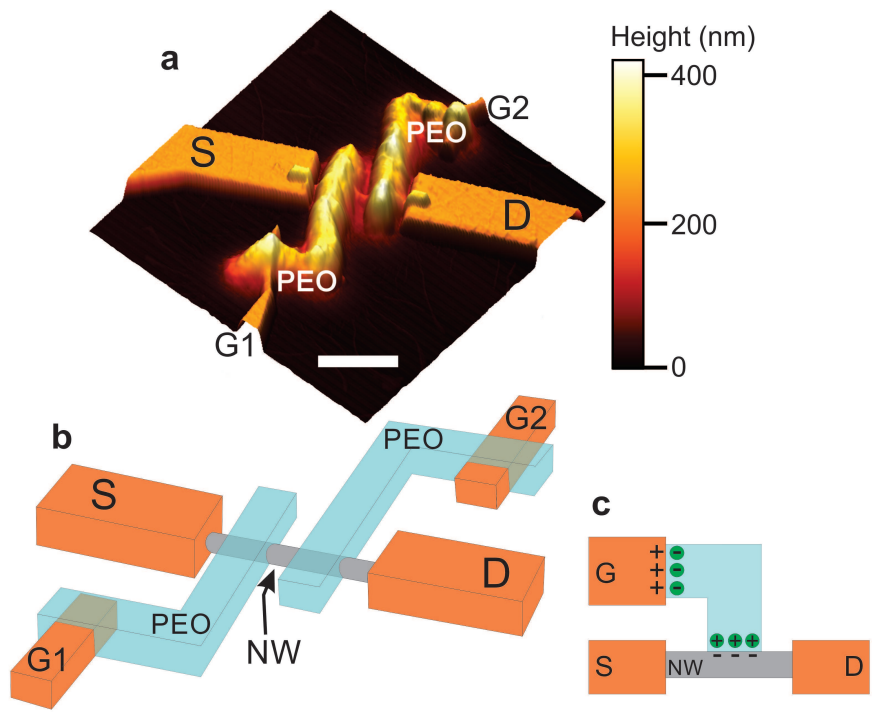}
\caption{\textbf{PEO-gated nanowire field-effect transistor.} \textbf{a} Atomic force micrograph and \textbf{b} schematic showing the $n$-InAs or $p$-GaAs nanowire (NW) channel with metal source (S) and drain (D) contacts and gate electrodes (G1 and G2). The gates and nanowire were linked by electron-beam patterned PEO.~\cite{CarradNL14} The schematic in c) illustrates the separation of ionic species, e.g., H$^+$/OH$^-$, Li$^+$/ClO$_4^-$ across the PEO in response to an applied gate voltage. G1 and G2 were offset significantly from the channel here to ensure gating only occurs by proton drift and not by direct field-effect (see Supplementary Fig.~S1).~\cite{ChoNatMat08} The scale bar in \textbf{a} represents $2~\mu$m.}
\end{figure}

Here we report a new hybrid organic-inorganic materials platform with considerable scaling potential that presents an interesting route to overcoming these problems. We combine salt-free nanoscale polyethylene oxide (PEO) strips patterned by electron-beam lithography with $n$-type InAs and $p$-type GaAs nanowires to demonstrate proton-to-electron transduction at frequencies up to $50$~Hz. This builds on previous work on polymer electrolyte-gated nanowire devices~\cite{LiangNL12, CarradNL14} but improves their biocompatibility by removing the commonly added perchlorate salt. We also present functional prototype hybrid protonic/electronic complementary circuits, a first step towards direct biological signal amplification and bioelectronic logic.~\cite{StrakosasJAPS15, TybrandtNatComm12} We attribute the transduction efficacy of our devices to the high surface-to-volume ratio of both the nanowire conducting channel and the patterned PEO dielectric, combined with the low area interface between them. In particular the high surface-to-volume ratio of our nanoscale PEO strips should drive significantly enhanced H$_{2}$O uptake compared to the thin-film parallel-plate capacitor geometries previously used to study ionic conductivity in PEO. This translated to a drastic improvement in PEO ionic conductivity and electrical performance. Strong ion-to-electron transduction can be obtained in inorganic semiconductor nanowires by virtue of their high surface-to-volume ratio and strongly surface-influenced conduction properties.~\cite{CuiSci01,MisraPNAS09} These desirable properties have seen individual nanowires both with and without functionalized surfaces used to detect specific molecules~\cite{CuiSci01, MisraPNAS09} and act as, e.g., gas~\cite{DuNL09} and pH sensors.~\cite{CuiSci01, UpadhyayAPL14} In our devices ionic separation across the PEO strip replaces polarization of the surface charge with the advantage that the ionic signal could potentially be simultaneously sensed by complementary transistors and thereby enable functional logic and amplification circuits. Additionally, by using an inorganic semiconductor channel we can tap into decades of materials optimisation and nanoscale integration knowledge enabling us to tailor our $n$- and $p$-type devices to the common operating voltage range necessary for functional complementary architectures. We chose III-V semiconductor nanowires over Si for their higher electron/hole mobility, noting also recent rapid improvements in direct integration of III-V nanowires with Si-based microelectronics.~\cite{TomiokaNat12, SchmidAPL15}

\section{Results}
\textbf{Device Design.} Figure~1a/b shows the nanowire transistor structure used in these experiments. Two electron-beam patterned PEO strips~\cite{KrskoLangmuir03, CarradNL14} extend from the end of metal gate electrodes (G1/G2) to cross an $n$-InAs or $p$-GaAs nanowire connected to source (S) and drain (D) contacts. Although in principle only one PEO strip is required to achieve a working device, our devices feature two for improved yield, to confirm proper strip isolation by independent gating,~\cite{CarradNL14} and to check reproducibility of performance. A gate voltage $V_{G}$ on G1 or G2 drives ion drift within the PEO leading to electric double layers at the interfaces between the PEO and gate electrode and the PEO and nanowire as shown schematically in Fig.~1c. These electrical double layers effectively transfer the gate charge to within $\sim 1$~nm of the nanowire surface, as discussed by Kim {\it et al.}~\cite{KimAdvMat13}, giving rise to a large gate capacitance and significantly reduced switching voltage. Typically the PEO is intentionally doped with a salt e.g. LiClO$_{4}$, and Li$^{+}$/ClO$_{4}^{-}$ transport is assumed to dominate the gating effect. The lack of salt-doping in our PEO and the device behaviour explored below instead point to H$^+$/OH$^-$ as the active ionic species, consistent with the known protonic conductivity behaviour of undoped PEO.~\cite{BinksJPolyPhys68} There H$^+$/OH$^-$ ions are transported along PEO chains via the ethereal oxygen atoms exploiting local segmental motions of the polymer. We would also expect the Grotthuss mechanism of proton transfer along hydrogen-bonded networks in the adsorbed water to play a role, as reported recently for other proton-conductive polymers.~\cite{ZhongNatComm11, DengSciRep13} Gating action at the nanowire surface will be a mixture of proton-accumulation field-effect and protonation/deprotonation of the nanowire's native surface oxide, as occurs in InAs nanowire ion-sensitive field-effect transistors (ISFETs).~\cite{CuiSci01,MisraPNAS09,UpadhyayAPL14} The gate electrode terminates $\sim4~\mu$m from the nanowire, alongside the source or drain contact. This `offset' gate geometry eliminates any competing influence by direct `vacuum' field-effect or non-ionic dielectric polarisation.~\cite{ChoNatMat08, CarradNL14} We do this to confirm ion transport is the dominant gating mechanism and to facilitate measurement of the PEO protonic conductivity; a detailed discussion is available in the Supplementary Information.

\begin{figure}
\includegraphics[width=12cm]{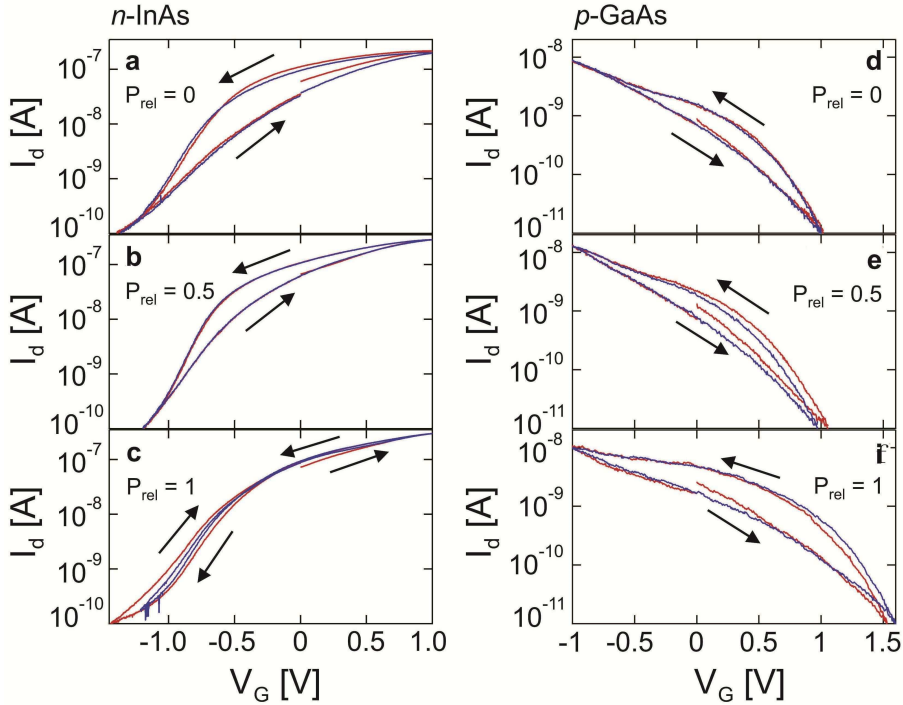}
\caption{\textbf{Transfer characteristics for III-V nanowire transistors with PEO gate dielectrics.} Plots of nanowire channel current $I_{d}$ vs polymer electrolyte gate voltage $V_{G}$ for \textbf{a-c} an InAs nanowire transistor and \textbf{d-f} a GaAs nanowire transistor, both with undoped PEO gate dielectric. In each case we present two consecutive gate traces (red/blue) obtained at three different relative water vapour pressures $P_{rel}$. Trace direction is indicated by arrows, with sweeps towards depletion terminated just prior to reaching the full `off' state to avoid the complex figure-of-eight pattern that emerges due to gate hysteresis.~\cite{CarradNL14} The lessening hysteresis with increased hydration is indicative of a proton-based gating mechanism. We present the entire data set in Supplementary Figs~S2 and S3, which allows us to comment further on the gate hysteresis and the non-monotonic behaviour for $p$-GaAs devices.}
\end{figure}

\textbf{Gating action and humidity dependence of undoped polyethylene oxide.} Figure~2 shows the gating action of our InAs and GaAs nanowire transistors with undoped PEO gate dielectrics. The InAs and GaAs devices in Fig.~2 exhibit clear $n$-type and $p$-type depletion, respectively. The devices are well matched, with subthreshold swings $\approx 300$~mV/dec, on/off ratios $\approx 10^3-10^4$ and well overlapping operating voltage ranges $-1.5~<~V_{G}~<~+1$~V, thereby showing significant potential for complementary circuit architectures. The lower current $I_{d}$ for the $p$-GaAs device on-state reflects the difficulty in making low resistance $p$-type ohmic contacts to GaAs nanowires; work continues towards improving this. We present data at three relative water vapour pressures $P_{rel} = 0$, $0.5$ and $1$ in Fig.~2; the full data set in $P_{rel}~=~0.125$ increments is in the Supplementary Information. Hydration control was achieved using a specially constructed vacuum chamber connected to a reservoir of degassed, deionized H$_2$O.~\cite{MostertLangmuir10, MostertAPL12} The existence of a hydration dependence is consistent with H$_2$O uptake, dissociation and H$^+$/OH$^-$ transport being the key component of the gating action in our devices. This behaviour is expected given the known high water uptake~\cite{KrskoLangmuir03} and evidence of proton conduction~\cite{BinksJPolyPhys68, DeLongchampLangmuir04} in PEO thin films (see Discussion for more details).

\begin{figure}
\includegraphics[width=12cm]{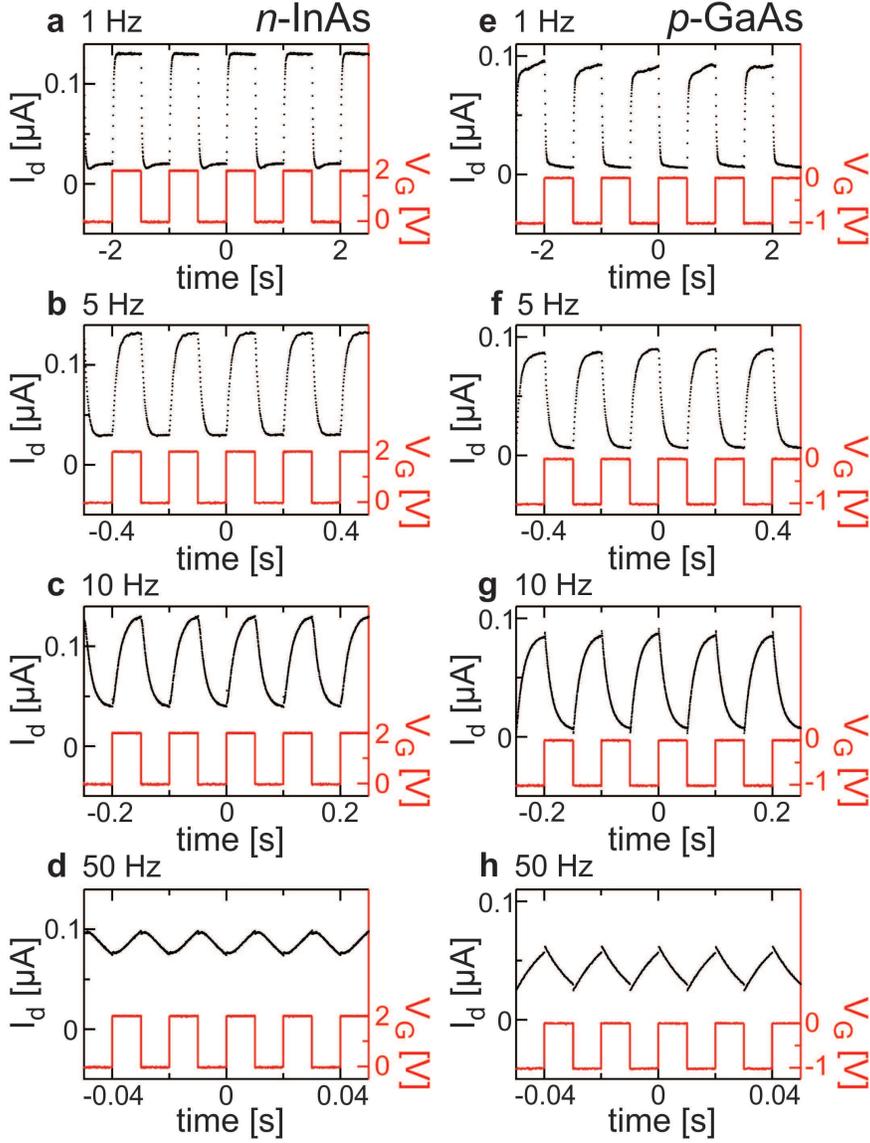}
\caption{\textbf{Switching performance of NWFETs with PEO gate dielectrics at $P_{rel} = 1$. a-h} Plots of nanowire channel current $I_d$ (black, left axis) vs time for a square wave applied to $V_G$ (red, right axis) with frequency $f = 1$, $5$, $10$ and $50$~Hz for \textbf{a-d} InAs and \textbf{e-h} GaAs nanowire channels. For this demonstration both devices were operated in a gate range where $I_d$ is linear in $V_G$ rather than the common range used in Fig.~4. The current response in this regime clearly shows the stability of both on and off states, and enables accurate determination of the protonic conductivity (see Supplementary information). Reduced fidelity at $f~>~10$~Hz is due limited proton conductivity and the $\sim 4~\mu$m length of the PEO gate dielectric strip.}
\end{figure}

The most strongly affected aspect of our devices under humidity increase was device operating speed. This manifested experimentally as hysteresis in the quasi-dc gate characteristics and frequency limiting under ac switching operations. Given the importance of operating speed for potential signal transduction applications, we now discuss the two key contributions of charge trapping by nanowire surface-states and the undoped PEO's proton conductivity. The gate hysteresis under quasi-dc conditions is substantially reduced at high $P_{rel}$ for the $n$-InAs devices (Fig.~2a-c). This indicates the importance of H$^+$/OH$^-$ transport in our devices and its increasing effect as H$_2$O adsorption and dissociation improves the proton conductivity. Some hysteresis remains at $P_{rel}~=~1$, characteristic of InAs surface-state trapping.~\cite{RoddaroAPL08} The $p$-GaAs devices (Fig.~2g-i) show more complex trends with increasing $P_{rel}$. Hysteresis initially decreases -- as for $n$-InAs devices -- but increases again for higher $P_{rel} > 0.75$ (the trends are more apparent in the full data set shown in Supplementary Fig.~S3). This likely indicates a stronger surface-state contribution for $p$-GaAs, consistent with literature for planar $p$-GaAs transistors.~\cite{BurkePRB12, CarradJPCM13} While the hysteresis is a feature of the quasi-dc characteristics, it has a lesser influence on the ac characteristics providing device operation is sufficiently fast that surface state trap occupancy remains quasi-static. One example is the GHz operation of InAs nanowire transistors~\cite{EgardNL10} even when the quasi-dc characteristics exhibit hysteresis.~\cite{LindNL06} We now turn to data showing that surface states do not impede high fidelity transduction of biological/ionic signals.

\begin{figure}
\includegraphics[width=12cm]{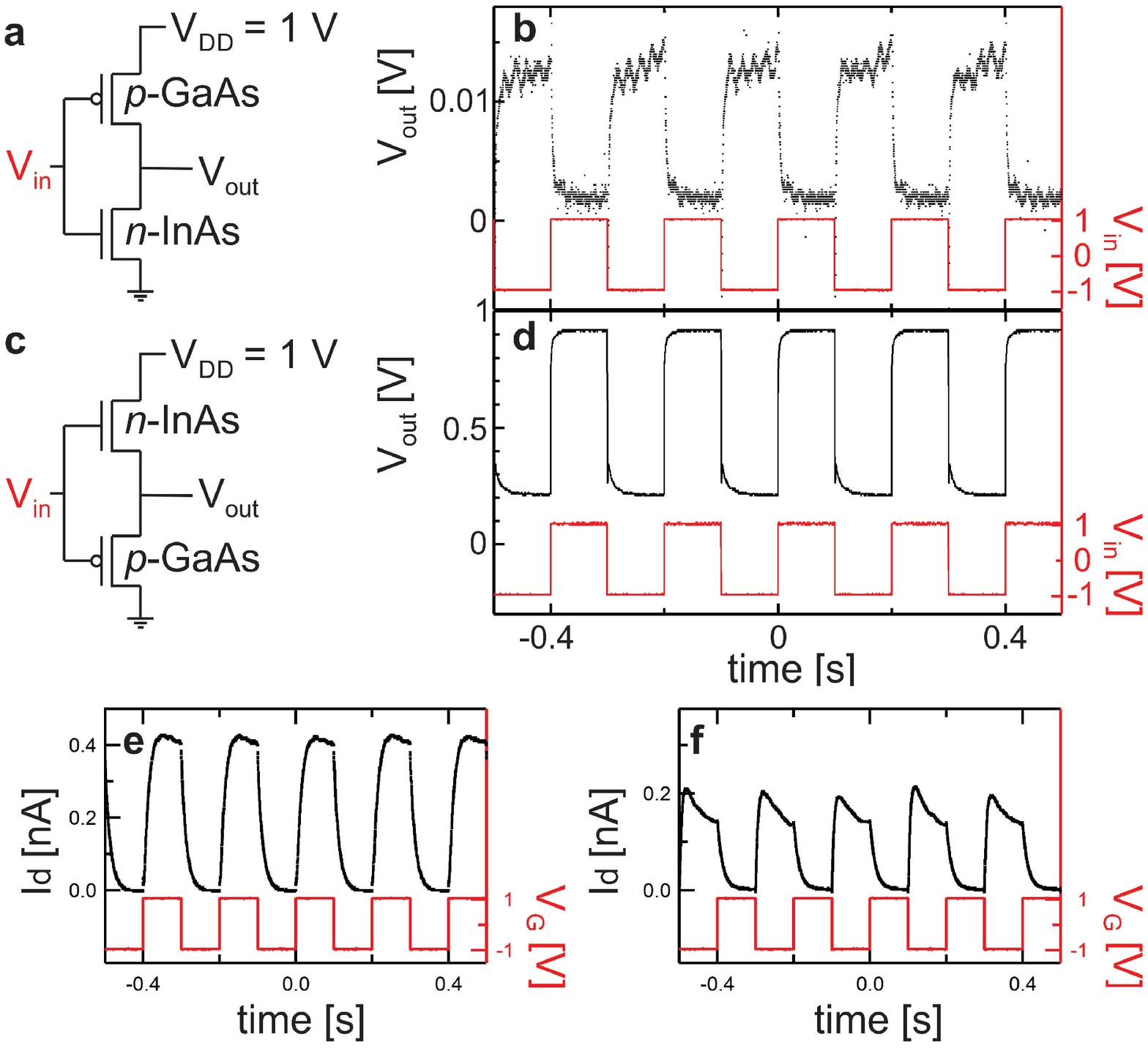}
\caption{\textbf{Demonstration of hybrid organic/inorganic complementary circuitry for logic and signal amplification} \textbf{a} Schematic of a traditional inverter circuit featuring undoped PEO nanowire transistors with $p$-GaAs and $n$-InAs channels. \textbf{b} Plot of input voltage $V_{in}$ (red, right axis) and output voltage $V_{out}$ (black, left axis) vs time for the inverter circuit in \textbf{a}. The operating frequency $f~=~5$~Hz and the supply voltage $V_{DD}~=~+1$~V. This circuit embodies the logical NOT operation. \textbf{c} similar circuit to \textbf{a} but with the $p$-GaAs and $n$-InAs transistors swapped to demonstrate the effect of high contact resistance in the $p$-GaAs device. \textbf{d} Plot of input voltage $V_{in}$ (red, right axis) and output voltage $V_{out}$ (black, left axis) vs time for the complementary circuit in \textbf{c}. The operating frequency $f~=~5$~Hz and the supply voltage $V_{DD}~=~+1$~V. Device optimisation could see this circuit used to achieve signal amplification. \textbf{e/f} Plots of channel current $I_d$ (black, left axis) and gate voltage $V_G$ (red, right axis) for the individual \textbf{e} $n$-InAs and \textbf{f} $p$-GaAs transistors used in \textbf{a/c}. Both are presented over the common $-1~<~V_G~<+1$~V operating range used in complementary circuit operation, as opposed to the linear regime used in Fig.~3.}
\end{figure}

\textbf{ac signal response.} The transistor response to ac gate signals that mimic biological signals is the most interesting aspect from a proton-to-electron transduction perspective. Figures~3a-h show the $I_d$ response (black/left axis) to a square wave $V_G$ input (red/right axis) for increasing frequency $f$ at $P_{rel}~=~1$. Each device is operated where $I_d$ is linear in $V_G$, giving good square wave fidelity at $f~=~1$~Hz (Fig.~3a/e). The slight $I_d$ noise/instability in the on-state arises from surface-state effects during the square-wave maxima/minima. Correspondingly the on-state stability improves at higher $f$ due to the reduced time available between $V_G$ switching events. The other notable feature in Fig.~3 is the gradual loss in square-wave fidelity with increased frequency, due to the PEO's limited proton conductivity. The loss of flat maxima/minima for $f~\geq~10$~Hz indicates that the time between pulses has dropped below the time required for complete charge separation across the PEO strip. This separation time depends on the PEO strip length but also the proton concentration $n$ and proton mobility $\mu$ (see Supplementary Information). Thus, our choice of long PEO strips here is deliberate -- it enables us to accurately measure proton conductivity $\sigma = ne\mu$, where $e$ is the proton charge. These strips could be shortened in future devices. We return to proton conductivity in more detail in the Discussion section below.

\textbf{Hybrid protonic-electronic complementary circuit architectures.} A major goal in bioelectronics is to build amplifiers and logic circuits featuring proton-to-electron transducers. We can achieve this using complementary circuit architectures given our approach provides both $n$- and $p$-channel transistors. Figure~4 demonstrates two functional circuits featuring $p$-GaAs and $n$-InAs transistors with undoped PEO gate dielectrics. We mimicked a wet biological environment by opening the H$_2$O reservoir to the chamber for $5-10$~mins -- much longer than the $10-20$~s required in previous experiments -- to facilitate H$_2$O condensation on the sample surface and complete hydration of the PEO films. Cross-linking of the PEO to the substrate during the electron-beam patterning process ensures the nanoscale strips remain stable when immersed in H$_2$O and PBS solutions for periods exceeding many hours.~\cite{KrskoLangmuir03, HongLangmuir04} The first circuit is the traditional inverter circuit in Fig.~4a. The transistor channels were connected in series with the $p$-GaAs drain at $V_{DD}~=~1$~V and the $n$-InAs source at ground. Here it is essential that the $n$-type and $p$-type transistors have strongly overlapping operating gate voltage ranges because the gate electrodes of both transistors must be fed a common electronic square-wave $V_{in}$ (Fig.~4b - red) for complementary operation. The individual $n$- and $p$-type transistor responses over the common input voltage range are presented in Figs.~4e/f. The electronic output voltage $V_{out}$ in Fig.~4b (black) was extracted between the transistors and found to be inverted relative to $V_{in}$, effectively embodying the logical NOT operation. The limited $V_{out}$ in Fig.~4b is due to the high $p$-GaAs contact resistance; $V_{out}$ ideally swings between $V_{DD}$ and ground. We demonstrate this using the circuit in Fig.~4c where the $p$-GaAs and $n$-InAs devices are reversed. Figure~4d shows $V_{out}$ (black trace) now reaching close to $V_{DD}$ when $V_{in} = 1$V (red trace). The high $p$-GaAs contact resistance prevents $V_{out}$ reaching ground. The signal fidelity is much better and although the gain is less than one, with some optimisation it should be possible to achieve amplification with this circuit (this configuration is commonly known as a complementary push-pull amplifier). We note here the integral role played by the nanoscale patterned, proton-conducting PEO in enabling detection circuits capable of performing logic and amplification. The key is that the PEO can ultimately perform the initial signal detection with nanoscale precision. Thereafter the signal is distributed via proton transport to arrive simultaneously at the $n$- and $p$-nanowires. The simultaneous signal arrival is vital to the behavior in Fig.~4; high-fidelity amplification and/or logic cannot be obtained if the signals arrive at different times. Using a biocompatible protonic conductor such as PEO with a well-defined path length as the link between the detection site and the two transistors ensures the circuit functions correctly. Additional steps and challenges towards monolithic integration of our transducing circuits are addressed in the Supplementary Information.

\section{Discussion}
\textbf{Protonic conductivity of nanoscale strips of undoped PEO.} Improving signal fidelity and switching speed can be done by optimising device geometry, increasing proton conductivity, or both. As an important step towards this, we measured the protonic conductivity by utilising the $I_d$ transient response of single devices since it is governed by proton drift and follows the exponential form expected for an RC circuit.~\cite{KimAdvMat13} Using our knowledge of the PEO strip dimensions and capacitance, we obtain a proton conductivity $\sigma~\sim~10^{-7}$~S/cm (see Supplementary Information for full details, fitting data and $\sigma$ versus $P_{rel}$). Interestingly, the proton conductivity for our undoped PEO films is two to three orders of magnitude higher than previously reported for undoped PEO thin films measured in a parallel-plate electrode geometry.~\cite{Fullerton-ShireyMacroMol09, BinksJPolyPhys68} We attribute this to two key differences between our samples and prior studies.

Firstly, the nanoscale patterned PEO in our device geometry is highly exposed, has a tiny volume and a very high surface-to-volume ratio. By contrast, the parallel-plate geometry used in previous works is well known to strongly inhibit water uptake because water can only access via the small exposed side-areas of the thin-film. This can lead to significant inaccuracies (under-measurement) in hydration-dependent studies.~\cite{MostertAPL12,MostertPNAS12} As a result, we expect much higher real H$^+$/OH$^-$ concentration for our devices compared to previous works -- and a corresponding increase in $\sigma$ -- under comparable atmospheric conditions. Note that $\sigma$ may also be influenced by increased $\mu$ arising from the swelling behaviour of PEO upon water uptake,~\cite{KrskoLangmuir03} however we believe our data is best explained by increasing $n$ with increased $P_{rel}$ as detailed in the Supplementary Information. The enhanced efficacy of water uptake for our nanoscale PEO strips is an advantage for our devices because improved protonic conductivity brings improved frequency response, irrespective of whether higher $\sigma$ comes about from increased $n$ or $\mu$. The more traditional capacitor geometry is favoured instead for relative ease of fabrication and the precision with which the geometric parameters can be obtained. Note however that geometric parameter uncertainty for our device is insufficient to account for the much higher conductivity we observe; we can measure the dimensions of our PEO strips accurately using atomic force microscopy.

Secondly, in prior work care was taken to exclude water during fabrication for accurate dry-state measurements by, e.g., purifying the as-bought chemicals, carefully dehydrating them and conducting processing in an N$_2$ environment. We took no such steps to remove residual H$_2$O or exclude H$_2$O from the processing, indeed, it is used as the developer solution in our PEO electron-beam lithography process.~\cite{CarradNL14} Taken together, the device geometry and increased exposure to H$_2$O -- whether at development or simply by exposing the films to ambient -- explains well the comparatively high H$^+$/OH$^-$ concentration in our PEO films. PEO has a known strong affinity for adsorbed H$_2$O, which is difficult to remove without taking the temperature well above the PEO melting point under high vacuum conditions.~\cite{BinksJPolyPhys68} We cannot use this approach as it destroys our nanopatterned strips. We find strong proton conduction is retained even if our sample is held at high vacuum for over $15$~hours at room temperature, consistent with earlier work by Binks \& Sharples.~\cite{BinksJPolyPhys68} This is advantageous from an applied perspective as it improves stability against environmental changes during device operation. The ability to exclude salt-doping without sustaining performance losses is highly advantageous in terms of biocompatibility and fabrication simplicity. A potential issue with the use of InAs and GaAs nanowires is toxicity due to leaching of their surface oxides, which can be unstable under biological conditions. This can be mitigated by simple surface treatments to stabilize the oxide~\cite{ChoNanoBio06, JewettAccChemRes12, ParkAdvHealthMat14} but coverage by PEO may also aid with this. Additionally, signal transduction via the nano-patterned PEO strip means we can physically separate the biological detection site from the nanowires, enabling encapsulation of the nanowire/PEO interface with, e.g., parylene, to ensure fully biocompatible circuit design.

We briefly comment on why we believe ionogenic impurities, e.g., trace salt, play no significant role in our undoped PEO. Firstly, supplier assays for our PEO and solvents have salt contamination at $<0.3~\%$, which we have confirmed via our own inductively coupled plasma mass spectrometry analyses. Secondly, we can support this by noting that equivalent devices featuring PEO films doped at $9-12.5\%$ level with LiClO$_{4}$, as is conventional for PEO-based polymer dielectrics,~\cite{GoreckiElectroActa92,PanzerAPL05,Fullerton-ShireyMacroMol09,LiangNL12,KimAdvMat13} give approximately equivalent electrical performance (see Supplementary Figure~S2/3 and associated discussion). For sake of argument, let us momentarily assume contamination by a non-Li salt, e.g., Na$^{+}$ or K$^{+}$. These ions {\it a priori} should have similar mobility as they are subject to the same chain relaxation transport mechanism~\cite{GoreckiElectroActa92} as H$^{+}$ or Li$^{+}$. In reality, the mobility would be slightly lower by virtue of the higher ionic mass/size. For our undoped PEO to give ionic conductivity/performance comparable to Li-doped PEO, without protonic transport playing any role, we would require a similar ion concentration, and thus salt contamination at $\sim 10~\%$ level -- clearly this is inconsistent with our chemical analysis. The earlier work by Binks \& Sharples arrived at an identical conclusion, namely that protonic transport and not ionogenic impurity effects explain the conductivity behaviour of undoped PEO with significant water content.~\cite{BinksJPolyPhys68}

\textbf{The path to improved frequency performance.} The data presented here highlight the potential for nanowires in amplifying and processing biological signals; they also suggest points for enhancing our design. A key limitation presently is the frequency response. Our knowledge of the proton conductivity allows us to quantify potential improvements on this aspect. Shortening the PEO strip would be the simplest action but increasing its width to cover as much of the nanowire as possible without touching the contacts would also help. If we assume the PEO strip length can be reduced to $l~\sim~100$~nm with $\sigma~=~10^{-7}$~S/cm, we would expect switching speeds exceeding $1.6$~kHz, more than sufficient for neural interfacing applications where signal frequencies are typically below $200$~Hz.~\cite{KhodagholyNatComm13} Note also that our devices inevitably involve a two-stage electron-to-proton-to-electron transduction process since an electrical voltage signal $V_G$ applied to the gate electrode drives H$^+$/OH$^-$ transport which in turn affects $I_d$ in the nanowire. Faster performance may be expected if using PEO-coated nanowires as a direct protonic sensing element. That is, $V_{in}$ could arise from direct protonic/ionic coupling of the PEO to a biological system of interest, e.g., a neuron,~\cite{KhodagholyNatComm13} biological nanopore,~\cite{NoyAdvMat11} or protonic transistor channel,~\cite{ZhongNatComm11, OrdinarioNatChem14} rather than the gate electrode used for our basic demonstration here. This will be a subject of future work. Further improvements in conductivity and frequency response may also be possible by doping the PEO with additional H$^+$,~\cite{HashmiJPhysD90} or by substituting the PEO for a material with higher protonic conductivity such as maleic chitosan.~\cite{ZhongNatComm11, DengSciRep13} A critical requirement and challenge for any such substitute material is demonstrating nanoscale patternability. As discussed above, the patternability of PEO is an essential feature of our design as it enables simultaneous sensing of the biological by the $n$- and $p$-type nanowire channels.

\textbf{Conclusion.} In conclusion, proton conduction in nanopatterned PEO films can be used for highly-effective gating of both $n$-type InAs and $p$-type GaAs nanowire transistors. The proton conductivity for our undoped PEO is $2-3$ orders of magnitude higher than previous reports,~\cite{Fullerton-ShireyMacroMol09} likely from differences in device architecture and processing. We demonstrated proton-to-electron signal transduction with good fidelity at frequencies up to $50$~Hz. Gains in operating speed can in future be achieved by reducing the PEO thickness or doping the PEO with excess H$^+$.~\cite{DeLongchampLangmuir04, RatnerChemRev88} We also demonstrated basic hybrid organic/inorganic complementary circuit architectures including a prototype inverter and repeater/push-pull amplifier circuit, although gain will require lower contact resistance to our $p$-GaAs nanowires. Our work provides a clear path towards high fidelity proton-to-electron transduction devices and circuits enabling functions, e.g., logic and amplification, using nanowire transistors featuring nanoscale proton conducting elements.

\section{Methods}
\textbf{Device Fabrication.} Nanowires with lengths $3 - 6~\mu$m and diameters $50$~nm or $100$~nm were grown by either vapour-liquid-solid MOCVD or MBE for InAs and GaAs, respectively. Surface accumulation of electrons in nominally undoped InAs nanowires results in $n$-type conduction without doping. The $p$-GaAs nanowires were Be-doped with acceptor concentration approximately $N_A = 1 \times 10^{18}$~cm$^{-3}$. The $N_A$ value is expected to closely match that for planar growth, although some variation may occur along the radial axis.~\cite{CasadeiAPL13} After transferring the nanowires to the measurement substrate, source (S) and drain (D) contacts and gate electrodes (G1 and G2) were defined by electron beam lithography using a polymethylmethacrylate (PMMA) resist. The contacts and electrodes consisted of $25/75$~nm Ni/Au for InAs and $200$~nm AuBe for GaAs deposited by thermal evaporation. For InAs, passivation in an aqueous (NH$_4$)$_2$S$_x$ solution immediately prior to metal deposition ensured ohmic contacts. For GaAs, the native oxide was removed by a 30~s HCl etch prior to metal deposition and post-deposition annealing at $300^{\circ}$C for 30~s caused the AuBe to diffuse into the GaAs and form ohmic contacts. Polymer electrolyte thin films were formed by dissolving dry PEO, spin coating the solution onto the sample, and evaporating the methanol by baking at 90$^{\circ}$C for 30 mins. This is similar to the process described in Refs.~\cite{LiangNL12, PanzerAPL05}. The thin films were patterned directly by electron beam lithography in a Raith 150-Two system using an accelerating voltage of $5$~kV, dose of $100 - 150$~$\mu$C/cm$^2$ and typical beam current $\sim 15$~pA. Extended details on electron beam patterning PEO appear in Ref.~\cite{CarradNL14}.

\textbf{Electrical Characterization.} Electrical measurements were performed under a hydration controlled atmosphere in a specially constructed vacuum chamber. The chamber was connected to a reservoir of highly purified water isolated by a bleed valve. The water was deionised Milli-Q water ($18$~M$\Omega$) that had been degassed with three freeze-pump-thaw cycles. Opening the bleed valve admitted water vapour to the chamber, increasing chamber pressure $P$ in discrete steps from vacuum ($P = 0$~mbar, $P_{rel} = 0$) to saturated vapour pressure ($P = 24$~mbar, $P_{rel} = 1$) as described previously~\cite{MostertLangmuir10, MostertPNAS12} and in detail in the Supplementary Information. A Keithley K2450 was used to supply $V_{sd} = 2$~mV ($n$-InAs) or $V_{sd} = 1$~V ($p$-GaAs) and measure $I_d$. To obtain the quasi-dc characteristics in Fig.~2 an additional Keithley K2400 was used to supply $V_G$ and monitor the gate leakage current, which remained below the noise floor ($<0.1$~nA) throughout the experiments. $V_G$ was incremented at $5$~mV/s. For ac measurements (Figs.~3 and 4) a H.P. 33120A function generator supplied the square wave voltage to the gate while $I_d$ was amplified by a Femto DLPCA-200 low noise current amplifier and recorded using an Agilent DSO-X 3024A digital oscilloscope. For the complementary circuits in Fig.~4b/d, $V_{out}$ was measured directly by the oscilloscope without external amplification. An extended discussion of these circuits and the methods used to measure proton conductivity is given in the Supplementary Information.

{\bf Supporting Information.} Full details on device fabrication, comparison of devices with LiClO$_{4}$-doped and undoped PEO, full humidity dependence data, full protonic conductivity data, and comments on transient response and issues related to full monolithic integration of complementary circuits are included in Supporting Information. This material is available free of charge via the Internet at http://pubs.acs.org.

{\bf Author Contributions.} DJC, ABM, PM and APM conceived and directed the project. DJC, RU and AMB fabricated devices. DJC and ABM performed measurements and analysed data. HJJ, HHT and CJ provided InAs nanowires. PK and JN provided GaAs nanowires. DJC and APM wrote the manuscript, assisted by ABM and PM. All authors commented on the manuscript.

\ack This work was funded by the Australian Research Council (ARC), the University of New South Wales, the University of Queensland, Danish National Research Foundation and the Innovation Fund. APM acknowledges an ARC Future Fellowship (FT0990285) and DJC acknowledges Australian Nanotechnology Network Short Term Visit support. PM is an ARC Discovery Outstanding Research Award Fellow and the work at UQ was funded under the ARC Discovery Program (DP140103653). The Centre for Organic Photonics and Electronics is a strategic initiative of the University of Queensland. We thank Helen Rutlidge for conducting the inductively coupled plasma mass spectrometry measurements. This work was performed in part using the NSW and ACT nodes of the Australian National Fabrication Facility (ANFF) and the Mark Wainwright Analytical Centre at UNSW.


\section*{References}

\begin{thebibliography}{99}

\bibitem{BerggrenAdvMat07} Berggren, M.; Richter-Dahlfors, A. {\it Adv. Mater.} {\bf 2007}, {\it 19}, 3201−
3213.

\bibitem{OwensMRS10} Owens, R. M.; Malliaras, G. G. {\it MRS Bull.} {\bf 2010}, {\it 35}, 449−456.

\bibitem{KhodagholyNatComm13} Khodagholy, D.; Doublet, T.; Quilichini, P.; Gurfinkel, M.;
Leleux, P.; Ghestem, A.; Ismailova, E.; Herv\'{e}́, T.; Sanaur, S.; Bernard,
C.; Malliaras, G. G. {\it Nat. Commun.} {\bf 2013}, {\it 4}, 1575.

\bibitem{ZhongNatComm11} Zhong, C.; Deng, Y.; Roudsari, A. F.; Kapetanovic, A.; Anantram,
M.; Rolandi, M. {\it Nat. Commun.} {\bf 2011}, {\it 2}, 476.

\bibitem{OrdinarioNatChem14} Ordinario, D. C.; Phan, L.; Walkup, W. G., IV; Jocson, J. M.;
Karshalev, E.; H\"{u}̈sken, N.; Gorodetsky, A. A. {\it Nat. Chem.} {\bf 2014}, {\it 6},
596−602.

\bibitem{StrakosasJAPS15} Strakosas, X.; Bongo, M.; Owens, R. M. {\it J. Appl. Polym. Sci.} {\bf 2015},
{\it 132}, 41735.

\bibitem{TybrandtPNAS10} Tybrandt, K.; Larsson, K. C.; Richter-Dahlfors, A.; Berggren, M.
{\it Proc. Natl. Acad. Sci. U. S. A.} {\bf 2010}, {\it 107}, 9929−9932.

\bibitem{TybrandtNatComm12} Tybrandt, K.; Forchheimer, R.; Berggren, M. {\it Nat. Commun.}
{\bf 2012}, {\it 3}, 871.

\bibitem{DengSciRep13} Deng, Y.; Josberger, E.; Jin, J.; Roudsari, A. F.; Helms, B. A.;
Zhong, C.; Anantram, M. P.; Rolandi, M. {\it Sci. Rep.} {\bf 2013}, {\it 3}, 2481.

\bibitem{CarradNL14} Carrad, D. J.; Burke, A. M.; Lyttleton, R. W.; Joyce, H. J.; Tan,
H. H.; Jagadish, C.; Storm, K.; Linke, H.; Samuelson, L.; Micolich, A.
P. {\it Nano Lett.} {\bf 2014}, {\it 14}, 94−100.

\bibitem{ChoNatMat08} Cho, J. H.; Lee, J.; Xia, Y.; Kim, B.; He, Y.; Renn, M. J.; Lodge,
T. P.; Frisbie, C. D. {\it Nat. Mater.} {\bf 2008}, {\it 7}, 900−906.

\bibitem{LiangNL12} Liang, D.; Gao, X. P. A. {\it Nano Lett.} {\bf 2012}, {\it 12}, 3263−3267.

\bibitem{CuiSci01} Cui, Y.; Wei, Q.; Park, H.; Lieber, C. M. {\it Science} {\bf 2001}, {\it 293},
1289−1292.

\bibitem{MisraPNAS09} Misra, N.; Martinez, J. A.; Huang, S. J.; Wang, Y.; Stroeve, P.;
Grigoropoulos, C. P.; Noy, A. {\it Proc. Natl. Acad. Sci. U. S. A.} {\bf 2009}, {\it 106},
13780−13784.

\bibitem{DuNL09} Du, J.; Liang, D.; Tang, H.; Gao, X. P. {\it Nano Lett.} {\bf 2009}, {\it 9},
4348−4351.

\bibitem{UpadhyayAPL14} Upadhyay, S.; Frederiksen, R.; Lloret, N.; De Vico, L.;
Krogstrup, P.; Jensen, J. H.; Martinez, K. L.; Nyg{\aa}rd, J. {\it Appl. Phys.
Lett.} {\bf 2014}, {\it 104}, 203504.

\bibitem{TomiokaNat12} Tomioka, K.; Yoshimura, M.; Fukui, T. {\it Nature} {\bf 2012}, {\it 488}, 189−
192.

\bibitem{SchmidAPL15} Schmid, H.; Borg, M.; Moselund, K.; Gignac, L.; Breslin, C. M.;
Bruley, J.; Cutaia, D.; Riel, H. {\it Appl. Phys. Lett.} {\bf 2015}, {\it 106}, 233101.

\bibitem{KrskoLangmuir03} Krsko, P.; Sukhishvili, S.; Mansfield, M.; Clancy, R.; Libera, M.
{\it Langmuir} {\bf 2003}, {\it 19}, 5618−5625.

\bibitem{KimAdvMat13} Kim, S. H.; Hong, K.; Xie, W.; Lee, K. H.; Zhang, S.; Lodge, T.
P.; Frisbie, C. D. {\it Adv. Mater.} {\bf 2013}, {\it 25}, 1822−1846.

\bibitem{BinksJPolyPhys68} Binks, A. E.; Sharples, A. {\it Journal of Polymer Science Part A-2:
Polymer Physics} {\bf 1968}, {\it 6}, 407−420.

\bibitem{MostertLangmuir10} Mostert, A. B.; Davy, K. J. P.; Ruggles, J. L.; Powell, B. J.; Gentle,
I. R.; Meredith, P. {\it Langmuir} {\bf 2010}, {\it 26}, 412−416.

\bibitem{MostertAPL12} Mostert, A. B.; Powell, B. J.; Gentle, I. R.; Meredith, P. {\it Appl.
Phys. Lett.} {\bf 2012}, {\it 100}, 093701−093701−3.

\bibitem{DeLongchampLangmuir04} DeLongchamp, D. M.; Hammond, P. T. {\it Langmuir} {\bf 2004}, {\it 20},
5403−5411.

\bibitem{RoddaroAPL08} Roddaro, S.; Nilsson, K.; Astromskas, G.; Samuelson, L.;
Wernersson, L.; Karlstr\"{o}m, O.; Wacker, A. {\it Appl. Phys. Lett.} {\bf 2008}, {\it 92},
253509.

\bibitem{BurkePRB12} Burke, A. M.; Waddington, D. E. J.; Carrad, D. J.; Lyttleton, R.
W.; Tan, H. H.; Reece, P. J.; Klochan, O.; Hamilton, A. R.; Rai, A.;
Reuter, D.; Wieck, A. D.; Micolich, A. P. {\it Phys. Rev. B: Condens. Matter
Mater. Phys.} {\bf 2012}, {\it 86}, 165309.

\bibitem{CarradJPCM13} Carrad, D. J.; Burke, A. M.; Reece, P. J.; Lyttleton, R. W.;
Waddington, D. E. J.; Rai, A.; Reuter, D.; Wieck, A. D.; Micolich, A. P.
{\it J. Phys.: Condens. Matter} {\bf 2013}, {\it 25}, 325304.

\bibitem{EgardNL10} Egard, M.; Johansson, S.; Johansson, A.; Persson, K.; Dey, A.
W.; Borg, B. M.; Thelander, C.; Wernersson, L.; Lind, E. {\it Nano Lett.}
{\bf 2010}, {\it 10}, 809−812.

\bibitem{LindNL06} Lind, E.; Persson, A. I.; Samuelson, L.; Wernersson, L. {\it Nano
Lett.} {\bf 2006}, {\it 6}, 1842−1846.

\bibitem{HongLangmuir04} Hong, Y.; Krsko, P.; Libera, M. {\it Langmuir} {\bf 2004}, {\it 20}, 11123−
11126.

\bibitem{Fullerton-ShireyMacroMol09} Fullerton-Shirey, S. K.; Maranas, J. K. {\it Macromolecules} {\bf 2009}, {\it 42},
2142−2156.

\bibitem{MostertPNAS12} Mostert, A. B.; Powell, B. J.; Pratt, F. L.; Hanson, G. R.; Sarna,
T.; Gentle, I. R.; Meredith, P. {\it Proc. Natl. Acad. Sci. U. S. A.} {\bf 2012}, {\it 109},
8943−8947.

\bibitem{ChoNanoBio06} Cho, Y.; Ivanisevic, A. {\it NanoBiotechnology} {\bf 2006}, {\it 2}, 51−59.

\bibitem{JewettAccChemRes12} Jewett, S. A.; Ivanisevic, A. {\it Acc. Chem. Res.} {\bf 2012}, {\it 45}, 1451−
1459.

\bibitem{ParkAdvHealthMat14} Park, G.; et al. {\it Adv. Healthcare Mater.} {\bf 2014}, {\it 3}, 515−525.

\bibitem{GoreckiElectroActa92} Gorecki, W.; Belorizky, E.; Berthier, C.; Donoso, P.; Armand,
M. {\it Electrochim. Acta} {\bf 1992}, {\it 37}, 1685−1687.

\bibitem{PanzerAPL05} Panzer, M. J.; Newman, C. R.; Frisbie, C. D. {\it Appl. Phys. Lett.}
{\bf 2005}, {\it 86}, 103503.

\bibitem{NoyAdvMat11} Noy, A. {\it Adv. Mater.} {\bf 2011}, {\it 23}, 807−820.

\bibitem{HashmiJPhysD90} Hashmi, S. A.; Kumar, A.; Maurya, K. K.; Chandra, S. {\it J. Phys. D:
Appl. Phys.} {\bf 1990}, {\it 23}, 1307.

\bibitem{RatnerChemRev88} Ratner, M. A.; Shriver, D. F. {\it Chem. Rev.} {\bf 1988}, {\it 88}, 109−124.

\bibitem{CasadeiAPL13} Casadei, A.; Krogstrup, P.; Heiss, M.; R\"{o}̈hr, J. A.; Colombo, C.;
Ruelle, T.; Upadhyay, S.; S{\o}rensen, C. B.; Nyg{\aa}rd, J.; Fontcuberta i
Morral, A. {\it Appl. Phys. Lett.} {\bf 2013}, {\it 102}, 013117.

\end{thebibliography}
\end{document}